# RF TEST RESULTS FROM CRYOMODULE 1 AT THE FERMILAB SRF BEAM TEST FACILITY*

E. Harms#, K. Carlson, B. Chase, E. Cullerton, A. Hocker, C. Jensen, P. Joireman, A. Klebaner, T. Kubicki, M. Kucera, A. Legan, J. Leibfritz, A. Martinez, M. McGee, S. Nagaitsev, O. Nezhevenko, D. Nicklaus, H. Pfeffer, Y. Pischalnikov, P. Prieto, J. Reid, W. Schappert, V. Tupikov, P. Varghese, Fermilab, Batavia, IL 60510, U.S.A.

J. Branlard, DESY, Hamburg, Germany

*Abstract*

Powered operation of Cryomodule 1 (CM-1) at the Fermilab SRF Beam Test Facility began in late 2010. Since then a series of tests first on the eight individual cavities and then the full cryomodule have been performed. We report on the results of these tests and lessons learned which will have an impact on future module testing at Fermilab.

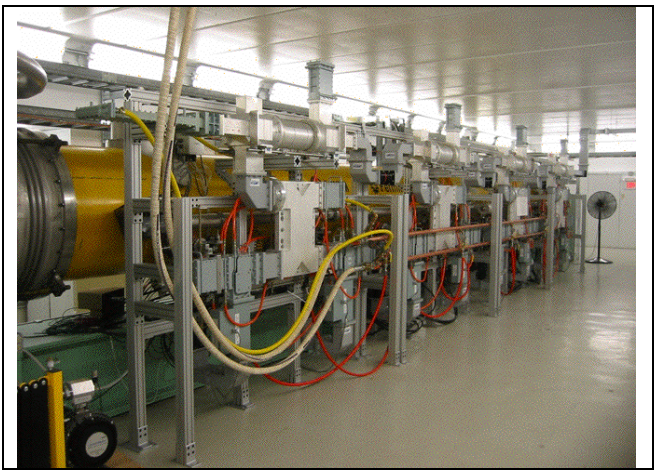

Figure 1: View of CM-1 from upstream to downstream with the waveguide distribution installed.

## INTRODUCTION

CM-1 is the first Tesla Type III Cryomodule to be placed into operation in the United States and the first multi-cavity cryomodule of any type to become operational at Fermilab. It was provided to Fermilab as a 'kit' from DESY in exchange for a 4-cavity 3.9 GHz module, ACC39, which was designed and assembled at Fermilab and is now in operation at DESY's FLASH facility. Assembly of the module has been documented previously [1]. Plans for a fully-utilized facility are described elsewhere at this conference. [2]. Table 1 highlights the major steps in bringing CM-1 into operation from its final installation in January 2010 to the present. Figure 1 shows CM-1 installed in the cave at the SRF Beam Test Facility.

Table 1: CM-1 Commissioning Milestones

| Milestone | Date |
|---|---|
| Cryomodule moved into final position and aligned | 22 January 2010 |
| Permission to initiate RF commissioning and warm coupler conditioning | 11 June 2010 |
| 5 MW RF/Klystron commissioning | June - July 2010 |
| Warm coupler conditioning - one cavity at a time (4 - 14 days/cavity) | August – October 2010 |
| Cryogenic and vacuum connections, leak check | October – November 2010 |
| Cool down from room temperature to 4 then 2 Kelvin | 17 – 22 November 2010 |
| Approval to initiate cold RF operation | 10 December 2010 |
| Cavity 1/Z89 cold conditioning and evaluation | 17 December 2010 - 26 January 2011, 18-22 March 2011 |
| Cavity 8/S33 cold conditioning and evaluation | 28 January 2011 - 7 March 2011 |
| Cavity 2/AC75 cold conditioning and evaluatio | 7 - 16 March 2011 |
| Cavity 3/AC73 cold conditioning and evaluation | 26 March - 4 April 2011 |
| Cavity 4/Z106 cold conditioning and evaluation | 20 April - 19 May 2011 |
| Cavity 5/Z107 cold conditioning and evaluation | 20 - 25 May |
| Cavity 6/Z98 cold conditioning and evaluation | 3 - 9 June |
| Cavity 7/Z91 cold conditioning and evaluation | 9 - 11 June |
| Installation of Waveguide Distribution system and Water System upgrade | 13 June - 5 July 2011 |
| First powering of CM-1 | 6 July 2011 |
| Evaluation and Operation of CM-1 | 7 July - |

## SRF BEAM TEST FACILITY OVERVIEW

CM-1 operation is made possible by a helium refrigerator/liquifier providing 110 Watts of cooling, a 5 MW klystron and associated high voltage power supply. Vacuum system, equipment interlock, and a controls system round out the infrastructure necessary for CM-1 operation.

### Low Level RF (LLRF) SYSTEM

The 1.3 GHz forward, reflected, and cavity probe signals of the CM-1 module cavities are down converted to 13 MHz and sent to a FPGA based digital signal processor (MFC, or Multiple-cavity Field Controller). Three down-converters with 8 channels each are used to do the down-conversion. The forward, reflected, and cavity signal are down-converted on separate boards. The down-conversion boards have a channel-to-channel isolation better than 76 dB. The LLRF system has a dynamic range of 153 dB, with a maximum input power of +10 dBm.

The 13 MHZ IF signals that are sent to the FPGA signal processor (MFC) are first down-converted to baseband I and Q signals. A vector sum of all eight cavity signals are used to provide a proportional / integral feedback to control the amplitude and phase. The MFC also uses feed-forward to generate 13 MHz I and Q signals that are up-converted, on the same board as the down conversion, to 1.3GHz to provide the RF drive to the klystron.

Calibration of the LLRF system was done using a calibrated RF source, and driving the signal into the forward power LLRF receiver chain. Measured data from the waveguide directional coupler was then added into the calibration. The reverse power was matched to the power at the beginning of the pulse when the cavity looks like a short circuit. Then, using the calibrated forward power, the cavity gradient was calibrated using the equation below.

The LLRF parameters and signals are controlled and displayed on a Labview™ interface which is tied into Fermilab's ACNET controls system. The Labview™ interface allows simple access to all the LLRF parameters and also allows for data analysis, such as FFT's and loaded Q measurements. The measured phase stability of the LLRF system is 0.05 deg/deg F, and amplitude stability is 0.009 dB/deg F

$$Cavity\ Voltage = \sqrt{4 \cdot P_{fwd} \cdot R/Q \cdot Q_l} \cdot (1 - e^{-t/\tau})/L$$

where,

$P_{fwd} = Forward\ power\ (Watts)$

$R/Q = 1036$

$Q_l = 3e6$

$\tau = 735\ \mu s$

$L = 1.038\ meters$

$t = time\ of\ P_{fwd}\ measurement\ (\mu s)$

### Motor Control System

The Cryomodule motion control system has the ability to drive up to 24 DC stepper motors at a time.

With this system all 8 'internal' cavity frequency adjustments along with 8 'external' coupler adjustments and 8 'external' phase shifters can be made remotely.

A typical motion system for the CM-1 module consists of the following commercial and Fermilab-designed components:
- 1 – VME based Motorola 5500 Power PC.
- 1 – IP #IPUCD – Fermilab designed Universal clock decoder.
- 2 – IP (Industry Pack) #IP330A – 16 channel (ea.) A/D modules by ACROMAG
- 6 – IP modules #8601 – 4 channels (ea.) DC stepper motor controller by Hytec Electronics.
- 3 – 8 channel (ea.) Motor driver chassis by ACS (advanced Control Systems)- STEP/PAK.
- 3 – VME based IP Carrier boards.

All Hardware is integrated into the VME5500 Processor card and is controlled through Fermilab's control system known as 'ACNET'.

This system will have the ability to auto-tune all components using frequency feedback for a closed loop auto-tune feature, but is not yet implemented.

## COMMISSIONING ACTIVITIES

The prescription followed to commission CM-1 once it was cooled to 2 Kelvin closely mimics the sequence used at DESY's Cryo Module Test Bench: individual power-up and coupler conditioning followed by full module operation [3]. For initial operation the waveguide between the klystron and CM-1 was connected to one cavity at a time and manually disconnected and re-attached to the directional coupler of the next cavity to be evaluated.

- Document the frequency spectrum
- Adjust resonant frequency to 1.300 GHz,
- Adjust $Q_L = 3.0\ E6$,
- Perform signal calibrations with a low power (few kW Forward power) and full pulse – 700 μs fill time and 500 μs flattop
- Determine initial $E_{Acc}$ and $k_T$
- Perform on-resonance conditioning with pulse widths of 20, 50, 100, 200. Peak power of 1MW is the goal with pulse widths up to 200 μs.
- Determine peak cavity operating gradient and source of limitation with 1200 μs pulse/5 Hz repetition rate
- Evaluate onset and magnitude of any field emission and dark current.
- Measure LLRF calibrations and demonstrate closed loop (feedback on) operation.
- Perform LFDC scans and demonstrate same Determine $Q_0$ vs. $E_{Acc}$ by measuring the Dynamic Heat Load at gradients close to the performance limit.

## Individual Cavity Evaluation

Each of the eight cavities showed a unique signature when it was tested by itself – time to condition, peak gradient, Q0, limiting factor, onset and magnitude of filed emission, dark current, for example. Full analysis a work in progress. Table 2 summarizes the performance of each cavity when powered individually and the limiting factor. level probe, perhaps indicating helium boiling. These cavities could reach a higher gradient threshold if the flattop length was reduced. Cavities #1/Z89, #3/AC73, and #7/Z91specifically show such behavior, an example of which is seen in figure 2.

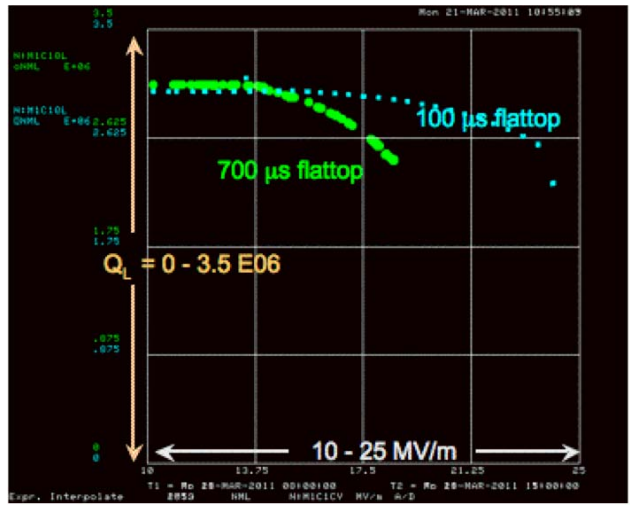

Figure 2: Variation of Cavity #1 $Q_L$ with Gradient, $E_{Acc}$. The $E_{Acc}$ plot limits are 10 to 25 MV/m. The green trace is with a 700 μs flattop and cyan with a 100 μs flattop.

'Soft' quenches are characterized by a drop in $Q_L$ and shortly thereafter by a response of the cryogenic system

These steps include:
evidenced by a change in the return pressure and liquid level if the gradient is increased.

While adjusting the resonant frequency for Cavity #8, the tuner motor ceased to operate after approximately twenty minutes of operation and a move of about -160 kHz from the as-found frequency. Troubleshooting has determined that the leads to the motor are shorted near the where they enter the motor housing. For this cavity's evaluation, the oscillator driving the LLRF system is set to put the cavity on resonance.

Table 2: CM-1 Cavity Performance Parameters

| Cavity # | Peak $E_{Acc}$ (MV/m) | Estimated Maximum $Q_0$ | Limitation/Comments |
|---|---|---|---|
| 1/Z89 | 18.5 | 1.1 E09 | 'Soft' quench/heat load |
| 2/AC75 | 21.5 | 1.2 E10 | Quench |
| 3/AC73 | 23 | 4.3 E08 | 'Soft' quench/heat load |
| 4/Z106 | 23.8 | 2.3 E09 | RF-limited |
| 5/Z107 | 27.5 | 3.9 E10 | Quench |
| 6/Z98 | 26 | 5.1 E09 | Quench |
| 7/Z91 | 22 | 4.7 E09 | 'Soft' quench/high heat load |
| 8/S33 | 25 | 1.8 E10 | Resonant frequency left at ~1300.240 MHz; tuner motor malfunction |

## Module Evaluation

Following completion of individual testing and a hiatus to upgrade the cooling water system necessary for sustained 5 HZ and high power operation of the RF system as well as install the waveguide distribution system, the entire module was powered. A vital step in installing the waveguide distribution system was adjusting the Variable Tap Offs (VTO's) which were set to match the outputs of a pair of adjacent cavities. Since initial powering additional checks on signal calibration have been made and are estimated to have an accuracy of 5% or better. With these final calibrations the maximum gradients as documented in table 2 and figure 3 were obtained. The values measured at DESY and CM-1 vary on average by ~20%. A combination of factors are suspected including calibration accuracy, venting and pump-down technique during assembly at Fermilab, as well as other factors yet to be determined. Typical operating conditions are:

- 5 Hz repetition rate
- 700 μs fill time
- 500 μs flattop
- 2 Kelvin / 25 Torr temperature
- 3.9 MW peak Klystron power.

The peak gradient was determined by evaluating the operability of the cavity measured – $Q_L$ no lower than 10% of the 3.0E6 value each cavity was set to, cryogenic stability, and hard quench limit as appropriate. LLRF closed loop operation has been demonstrated.

## REMAINING WORK

Recently LLRF closed loop operation has been demonstrated. Much work remains before testing is consider complete. Some tasks to complete and additional tests under consideration include:
- Demonstration of Lorentz Force Detuning Compensation on the entire module,
- Additional processing to attempt to improve the performance of deficient cavities,
- Possible thermal cycling to improve performance/uncover additional field emitters,
- Re-adjustment of VTO's to better match cavity pair outputs.

Study time is also envisioned for high $Q_L$ measurements related to Project X and LLRF and LFDC data gathering in support of ILC 9 mA studies.

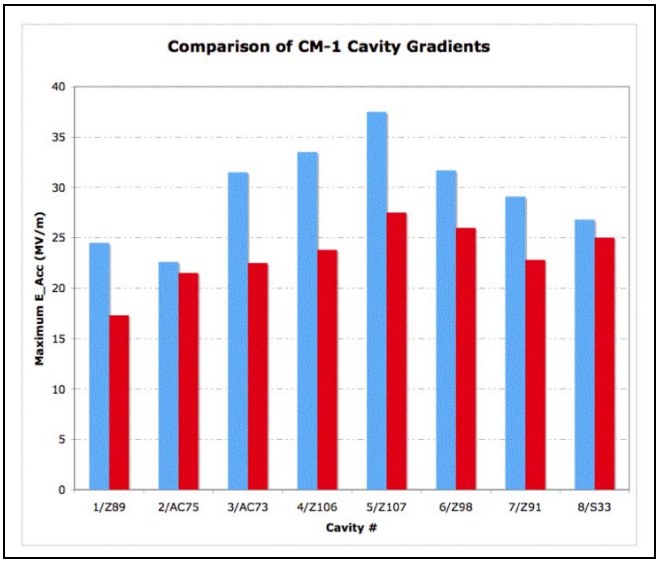

Figure 3: Comparison of peak Cavity gradients as measured at DESY's 'Chechia' horizontal test stand (cyan) and CM-1 at Fermilab (red).

## LESSONS LEARNED

Experience with learning to operate CM-1 in recent months is proving valuable in anticipation of bringing CM-2 and other cryomodules destined for this facility on line in the near future as well as in preparation for possible future SRF facilities such as Project-X. Already the suite of instrumentation has been reviewed and modified for CM-2 based on CM-1 experience. The sequence of commissioning steps have been modified during the this initial experience and will no doubt lead to faster and more efficient start-up of future modules.

In light of the tuner motor issues with Cavity #8, the motor controller circuitry was reviewed, modified, and a more robust and fail safe system is now realized.

The user interface system is also under regular updating as more experience is gained.

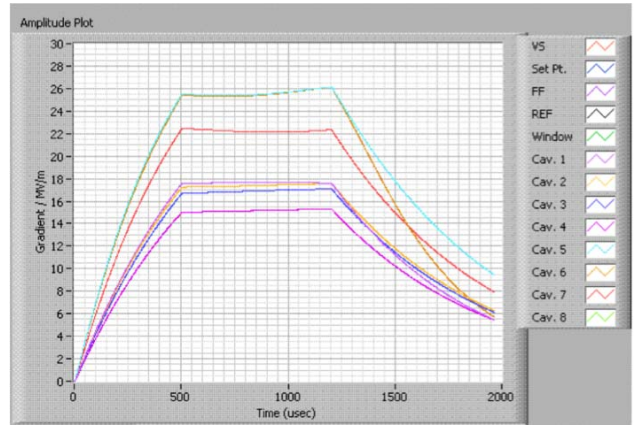

Figure 4: Low Level RF display of Cavity Amplitudes (MV/m) with all cavities powered simultaneously and VTO's adjusted for matched power to adjacent cavity pairs.

## SUMMARY

Since November 2010 Cryomodule 1 has been operating at 2 Kelvin. After evaluating each of the eight cavities while individually powered, the entire module has recently been powered and peak operation determined as shown in Figure 4. Several more weeks of measurements are planned before the module is warmed up, removed and replaced with Cryomodule 2 now under assembly at Fermilab.

## ACKNOWLEDGEMENTS

The achievements to date would not have been possible without the hard work and effort of many people in Fermilab's Technical and Accelerator Divisions. Shared expertise and contributions of hardware etc. from collaborators from many institutions around the world especially DESY, INFN, Milano, and SLAC are noted and appreciated.